\documentclass[11pt]{article}
\usepackage{amsmath}
\usepackage{amssymb}
\usepackage{wrapfig}
\usepackage{caption}
\usepackage{xcolor}
\usepackage[linesnumbered,boxed]{algorithm2e}
\usepackage{graphicx}
\usepackage{hyperref}
\usepackage[margin=1.25in]{geometry}

\newtheorem{theorem}{Theorem}[section] 
\newtheorem{lemma}[theorem]{Lemma}
\newtheorem{claim}[theorem]{Claim}

\newenvironment{proof}{\trivlist\item[]\emph{Proof}:}%
                  {\unskip\nobreak\hskip 1em plus 1fil\nobreak%
                           \rule{2mm}{2mm}
                           \parfillskip=0pt%
                           \endtrivlist}

\definecolor{blue25}{rgb}{0,0,0.25}
\newcommand{\emphic}[2]{%
     \textcolor{blue25}{%
         \textbf{\emph{#1}}}%
         \index{#2}}
\newcommand{\emphi}[1]{\emphic{#1}{#1}}

\newcommand{\lemlab}[1]{\label{lemma:#1}}
\newcommand{\lemref}[1]{Lemma~\ref{lemma:#1}}
\newcommand{\figlab}[1]{\label{fig:#1}}
\newcommand{\figref}[1]{Figure~\ref{fig:#1}}
\newcommand{\algolab}[1]{\label{algo:#1}}
\newcommand{\algoref}[1]{Algorithm~\ref{algo:#1}}
\newcommand{\seclab}[1]{\label{sec:#1}}
\newcommand{\secref}[1]{Section~\ref{sec:#1}}
\newcommand{\thmlab}[1]{\label{theorem:#1}}
\newcommand{\thmref}[1]{Theorem~\ref{theorem:#1}}
\newcommand{\apndxlab}[1]{\label{appendix:#1}}
\newcommand{\apndxref}[1]{Appendix~\ref{appendix:#1}}

\newcommand{\ccr}{\textrm{CanCover}^{\textrm{R}}}
\newcommand{\ccb}{\textrm{CanCover}^{\textrm{B}}}

\newcommand{\PntSet}{P}

\newcommand{\Mspace}{\mathcal{M}}
\newcommand{\pnt}{p}

\newcommand{\query}{\mathsf{q}}
\newcommand{\ball}[1]{\mathsf{B}(#1)}

\newcommand{\eps}{\varepsilon}
\renewcommand{\Re}{{\rm I\!\hspace{-0.025em} R}}

\newcommand{\abs}[1]{| #1 |}

\newcommand{\dist}[1]{\mathsf{dist}(#1)}
\newcommand{\Line}{\ell}
\newcommand{\Int}{I}
\newcommand{\IntSet}{\mathcal{I}}
\newcommand{\lept}{a}
\newcommand{\rept}{b}
\newcommand{\CandSet}{\mathcal{C}}
\newcommand{\cpt}{\mathsf{c}}
\newcommand{\cl}[1]{\bar{#1}}
\newcommand{\nxt}[1]{\mathsf{s}(#1)}
\newcommand{\ropt}{r^*}
\newcommand{\remove}[1]{}

\newcommand{\seq}{\mathsf{SEQ}}
\newcommand{\true}{\mathsf{\mathbf{True}}}
\newcommand{\false}{\mathsf{\mathbf{False}}}
\newcommand{\ceil}[1]{\lceil #1 \rceil}

\SetCommentSty{mycommfont}
\SetKwInput{KwInput}{Input}                
\SetKwInput{KwOutput}{Output}              

\title{Separated Red Blue Center Clustering}

\author{Marzieh Eskandari \thanks{Department of Computer Science, Alzahra University, Tehran, Iran \texttt{eskandari@alzahra.ac.ir}; \texttt{\url{https://staff.alzahra.ac.ir/eskandari/en/}}} 
       \and
       Bhavika B. Khare \thanks{Department of Computer Science, University of Memphis, Memphis, TN, USA \texttt{bbkhare@memphis.edu}}       
       \and
       Nirman Kumar \thanks{Department of Computer Science, University of Memphis, Memphis, TN, USA \texttt{nkumar8@memphis.edu}; \texttt{\url{https://www.cs.memphis.edu/\~ nkumar}}}
       }

\begin{document}

\maketitle

\begin{abstract}
We study a generalization of $k$-center clustering,
first introduced by Kavand et. al., where instead
of one set of centers, we have two types of centers, 
$p$ red and $q$ blue, and where each red center is 
at least $\alpha$ distant from each blue center. The goal is to minimize the covering radius. We provide
an approximation algorithm for this problem, and a 
polynomial time algorithm for the constrained
problem, where all the centers must lie on a line $\Line$.
\end{abstract}

\section{Introduction}\seclab{intro}
The $k$-center problem is a well known geometric location problem, where we are given 
a set $\PntSet$ of $n$ points in a metric space and a positive integer $k$, and the task is to
find $k$ balls of minimum radius whose union covers $\PntSet$. This problem can be used
to model the following facility location scenario. Suppose we want to open $k$ 
facilities (such as supermarkets) to serve the customers in a city. 
It is common to assume that a customer shops at the facility closest to their residence.
Thus, we want to locate $k$ locations to open the facilities, so that the maximum distance 
between a customer and their nearest facility is minimized. 
The problem was first shown to be
NP-hard by Megiddo and Supowit \cite{megiddosupotwik} for Euclidean spaces.
We consider a variation of this classic problem where instead of just one set of centers, we consider
two sets of centers, one of size $p$, and the other of size $q$, but with the constraint that
each center of the first set is separated by a distance of at least some given $\alpha$ from each
center of the second set. This follows from a more practical facility location scenario, where
we want to open two types of facilities (say `Costco's' and `Sam's club'). Each facility type wants
to cover all the customers within the minimum possible distance (similar to the $k$-center clustering
objective), but the facilities want to be separated from each other to avoid crowding or getting 
unfavorably affected by competition from the other.

The $k$-center problem has a long history. In 1857 Sylvester \cite{sylvester} presented the 1-center problem for the first time, and Megiddo \cite{megiddo} gave a linear time algorithm for solving this problem, also known as the minimum enclosing ball problem, in 1983, using 
linear programming. Hwang et al. \cite{hwang} showed that the Euclidean $k$-center problem in the plane can be solved in $n^{O(\sqrt{k})}$. Agarwal and Procopiuc \cite{agarwal} presented an $n^{O(k^{1-1/d})}$-time algorithm for solving the $k$-center problem in $\Re^d$ and a $(1 + \epsilon)$-approximation algorithm with running time $O(n \log k) + (k/ \epsilon)^{O(k^{1-1/d})}$.

Due to the importance of this problem, many researchers have considered 
variations of the basic problem to model different situations. Brass et al. studied the constrained version of the $k$-center problem in which the centers are constrained to be co-linear \cite{brass}, also considered previously
for $k=1$ by Megiddo \cite{megiddo}. They gave an  $O(n \log^ 2 n)$-time algorithm when the line is fixed in advance. Also, they solved the general case where the line has an arbitrary orientation in $O(n^4 \log^2 n)$ expected time. They presented an application of the constrained $k$-center in wireless network design: For a given set of $n$ sensors (which are modeled as points), we want to locate $k$ base servers (centers of balls) for receiving the signal from the sensors. The servers should be connected to a power line, so they have to lie on a straight line which models the power line. Other variations have also been
considered \cite{hurtado,boselangerman,bosetoussaint} for $k=1$. For $k \geq 2$, variants have been studied
as this has applications to placement of base stations
in wireless sensor networks \cite{das,roy,shin}. 

Hwang et al. \cite{hwang} studied a variant somewhat opposite to our variant. In their variant, for a given constant $0 \leq \alpha \leq 1$, the $\alpha$-connected two-center problem is to find two balls of minimum radius $r$ whose union covers the points, and the distance of the two centers is at most $2(1 - \alpha)r$, i.e., any two of those balls intersect such that each ball penetrates the other to a distance of at least $2 \alpha r$. They presented an $O(n^2 \log^2 n)$ expected-time algorithm. 

The variant we consider was first considered by Kavand et. al. \cite{kavand}. They termed it as the $(n,1,1,\alpha)$-center problem. 
Their aim was to find two balls each of which covers the entire
point set, the radius of the bigger one is minimized, and the distance of the two centers is at least $\alpha$. They presented an $O(n\log n)$-time algorithm for this problem, and a linear time algorithm for its constrained version using the furthest point Voronoi diagram. 

This paper considers the generalization of the problem defined by \cite{kavand}, and we denote it by $(n,p \land q,\alpha)$ problem. We explain our choice of notation, particularly the $\land$ sign, in \secref{defs}. For a given set $\PntSet$ of $n$ points in a metric space and integers $p, q \geq 1$, we want to find $p+q$ balls of two different types, called \emphi{red} and \emphi{blue} with the minimum radius such that $\PntSet$ is covered by the $p$ red balls and also covered by the second type of $q$ blue balls, and the distance of the centers of each red ball from the centers of the blue balls is at least $\alpha$.
In addition to one example mentioned before, another motivating application of the $(n,p \land q,\alpha)$ problem would be to locate $p$ police stations and $q$ hospitals in an area such that the distance between each police station and a hospital is not smaller than a predefined distance $\alpha$ for the convenience of patients.
By locating hospitals and police stations at an admissible distance from each other, patients stay away from crowd and noise while the clients have access to  hospitals and police stations which are close enough to them.
Moreover, it is obviously desirable that the maximum distance between a client and its nearest police station as well as its nearest hospital is minimized. 
In addition to this general problem we also consider the constrained version due to its applicability in many situations, where the centers are constrained to lie on a given line.

\noindent \textbf{Paper organization.} In \secref{defs}, we present the formal problem statement
and the definitions required in the sequel. In \secref{approx}, we present an $O(1)$ factor approximation
algorithm for the problem in Euclidean spaces. 
Then, in \secref{constrained-dp}, we present a
polynomial time algorithm for the constrained problem. We conclude in \secref{conclusions}.

\section{Problem and Definitions} \seclab{defs}
Let $\Mspace$ denote a metric space. 
Let $\dist{\pnt,\query}$ denote the distance between points $\pnt, \query$ in $\Mspace$.
For a point $x \in \Mspace$ and a number $r \geq 0$ the ball $\ball{x,r}$ is
the set of points with distance at most $r$ from $x$, i.e.,
$\ball{x,r} = \{ \pnt \in \Mspace | \dist{x,\pnt} \leq r \}$ is the \emph{closed} ball
of radius $r$ with center $x$.

In the $\alpha$-separated red-blue clustering problem we are given a set $\PntSet$ with $n$ points in some metric space $\Mspace$, integers $p > 0, q > 0$, and a real number $\alpha > 0$. For a given number $r \geq 0$, $p$ points $c_1, \ldots, c_p$ in $\Mspace$ (with possibly repeating points) called the red centers and $q$ points $d_1, \ldots, d_q$ in $\Mspace$ (with possibly repeating points) called the blue centers are said to be a \emphi{feasible solution} for the problem, with \emphi{radius
of covering} $r$ if they satisfy,
\begin{itemize}
\item \textbf{Covering constraints:}  The union of the balls 
$\bigcup_{i=1}^p \ball{c_i,r}$ (called
the red balls) covers $\PntSet$, and the union of the balls $\bigcup_{j=1}^q \ball{d_j, r}$ (called the blue balls) covers $\PntSet$. 

\item \textbf{Separation constraint:} For each $1 \leq i \leq p, 1 \leq j \leq q$ we have $\dist{c_i, d_j} \geq \alpha$, i.e., the red and blue centers are separated by at least a distance of $\alpha$.
\end{itemize}
If there exists a feasible solution for a certain value of
$r$, such an $r$ is said to be feasible for the problem. The goal of the problem is to find the minimum possible value of $r$ that is feasible.

We denote this problem as the $(n,p \land q, \alpha)$-problem. The $\land$ in the
notation stresses the fact that \emph{both} the red balls \emph{and} the blue balls cover $\PntSet$. 
Let $r_{p \land q,\alpha}(\PntSet)$ denote the optimal radius for this problem. When $\PntSet, p, q, \alpha$ are clear from context sometimes we will also denote this by $\ropt$. Also, let $r_{k}(\PntSet)$ denote the optimal $k$-center clustering radius, for all $k \geq 1$. To be clear,
the centers in the $k$-center clustering problem can be
any points in $\Mspace$, not necessarily belonging to
$\PntSet$. If that is the requirement, the problem is 
the \emph{discrete} $k$-center clustering problem.

For this paper, we will always be concerned with 
$\Mspace = \Re^d$, but we will use the notations as defined above without qualifying the metric space. We let $\PntSet = \{\pnt_1, \ldots, \pnt_n\}$ where,
$\pnt_i = (\pnt_{i1}, \pnt_{i2}, \ldots, \pnt_{id})$. We also consider the \emph{constrained} $\alpha$-separated red-blue clustering problem (when $\Mspace = \Re^d$) we are given
a line $\ell$ and all the red and blue centers are constrained to lie on the
line $\ell$. Without loss of generality, we will assume that $\ell$ is the $x$-axis since
this can be achieved by an appropriate affine transformation of space. Moreover, we will
use the same notation for the optimal radii and centers. For the constrained
problem we need some additional definitions and notations.  
We assume that no two points in $\PntSet$ have the same distance from $\ell$. (This general position assumption can however be removed.)
For each point $\pnt_i$, we consider the set of points on the line ($x$-axis) such that the
ball of radius $r$ centered at one of those points can cover $\pnt_i$. This
is the intersection of $\ball{\pnt_i, r}$ with the $x$-axis, see \figref{circ}.
Assuming this intersection is not empty, let the interval be 
$\Int_i(r) = [\lept_i(r), \rept_i(r)]$. Denote the set
of all intervals as $\IntSet(r) = \{ \Int_1(r), \ldots, \Int_n(r) \}$ where we 
assume that the numbering is in the sorted order of intervals: those
with earlier left endpoints are before, and for the same left endpoints
the ones with earlier right endpoint occurs earlier in the order. Notice that feasibility of radius $r$ means that there
exist two hitting sets for the set of intervals $\IntSet(r)$,
the red centers and the blue centers such that they satisfy the separation constraint.
\setlength{\intextsep}{-3pt}%
\begin{wrapfigure}{r}{0.5\textwidth}
  \centering
  \captionsetup{justification=centering}
    \includegraphics[width=0.50\textwidth]{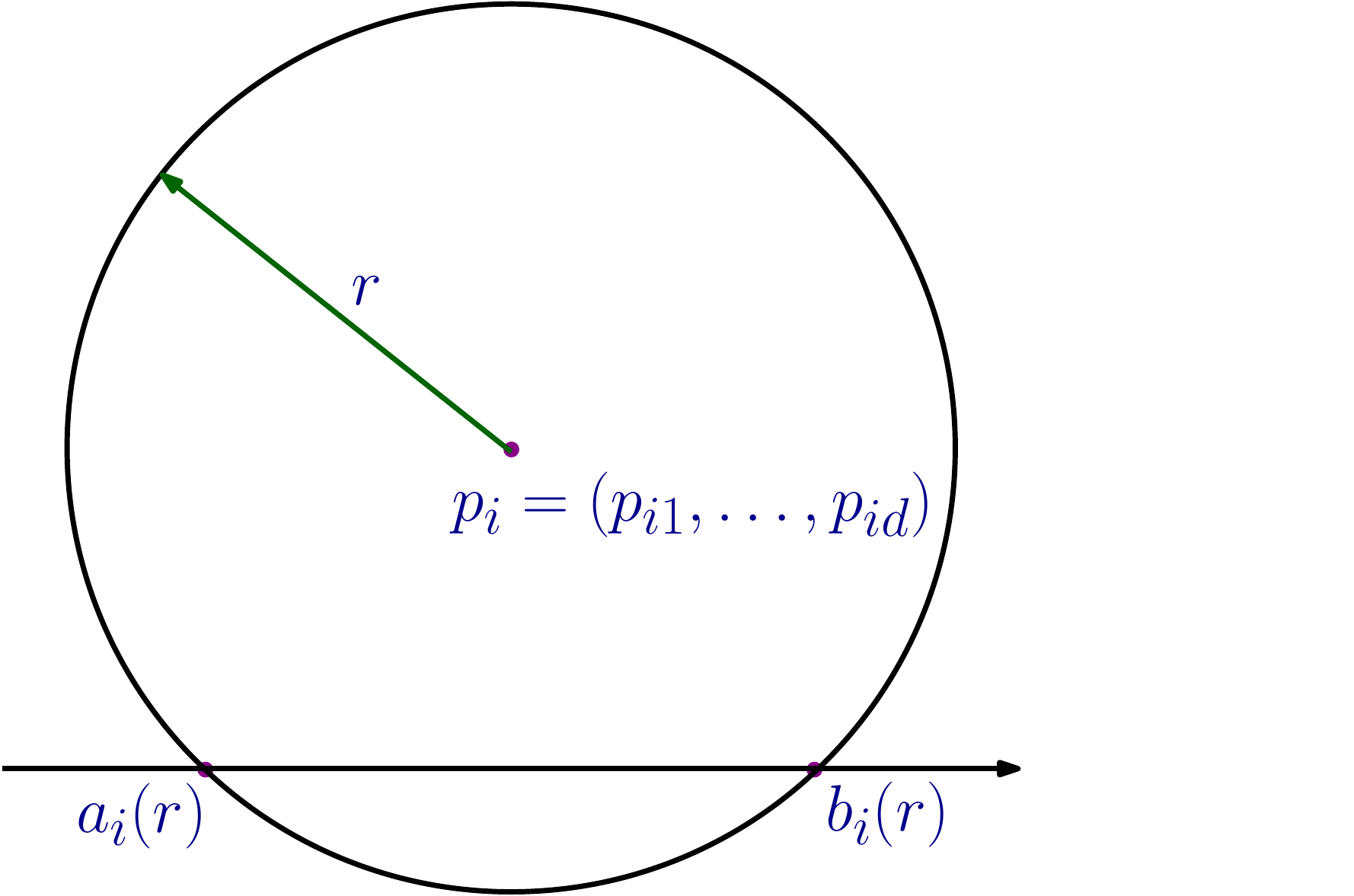}
    \caption{The functions $\lept_i(r), \rept_i(r)$}
    \figlab{circ}%
\end{wrapfigure}

The interval
endpoints $\lept_i(r), \rept_i(r)$ can be computed by solving the equation,
$$ (x-\pnt_{i1})^2 + \pnt_{i2}^2 + \ldots + \pnt_{id}^2 = r^2,$$ for $x$. Thus, they
are given by $\lept_i(r) = p_{i1} - \sqrt{r^2 - \sum_{j=2}^d p_{ij}^2}$,
and $\rept_i(r) = p_{i1} + \sqrt{r^2 - \sum_{j=2}^d p_{ij}^2}$. It is easy to see
that for the range of $r$ where the intersection is non-empty, $\lept_i(r)$ is 
a strictly decreasing function of $r$ and $\rept_i(r)$ is a strictly increasing function
of $r$.

\noindent \textbf{Model of computation.} We remark that our model of computation is the Real RAM model, where the usual arithmetic operations are assumed to take $O(1)$ time.

\section{Approximation algorithms for the \texorpdfstring{$(n, p \land q, \alpha)$}{npqa} problem in \texorpdfstring{$\Re^d$}{Red}} \seclab{approx}
The $(n,p \land q, \alpha)$ problem is NP-hard when $p, q$ are part of the input, since the
$k$ center problem clearly reduces to the $(n, p \land q, \alpha)$ problem when $\alpha = 0$ and 
$p + q = k$. Here we show 
an approximation algorithm for the problem as well as one with a better approximation factor for the constrained problem.

\subsection{Approximation for the general case}
\seclab{general-approx}%

Here we show that there is a constant factor algorithm
for the $(n,p \land q, \alpha)$ problem in $\Re^d$. We 
need a few preliminary results.
\begin{lemma}
	\lemlab{lem1}%
Suppose that $r \geq \alpha/2$ is a number such that there are points $x_1, \ldots, x_p$ satisfying
	$\PntSet \subseteq \bigcup_{i=1}^p \ball{x_i,r}$. Then, there are
	$(p+q)$ points $c_1, \ldots, c_p, d_1, \ldots, d_q$ all such that the following are met,
	(I) \emph{Separation constraint: } $\dist{c_i, d_j} \geq \alpha$ for all $i,j$, and,

	(II) \emph{Covering constraints: } $\PntSet \subseteq \bigcup_{i=1}^p \ball{c_i,7r}$, and, $\PntSet \subseteq \bigcup_{j=1}^q \ball{d_j,7r}$.
\end{lemma}
\begin{proof}
	Let $p \leq q$, wlog.
	First, from the points $x_1, \ldots, x_p$ we choose a maximal subset of them 
	such that the distance between each pair of them is at least $4r$. This can be
	done by a simple scooping algorithm that starts with $x_1$ as first point, then
	throws away all points $x_i$ (for $i > 1$) with $\dist{x_1, x_i} < 4r$, 
	then choose any one of the remaining points and proceed analogously.

	Suppose after this (with some renaming) the points that remain are,
	$x_1, \ldots, x_t$, where $1 \leq t \leq p$. Then, one can easily show that, $\PntSet \subseteq \bigcup_{i=1}^t \ball{x_i, 5r}$.

    Now, we choose the $p$ red centers $c_1, \ldots, c_p$ at the
    points $x_1, \ldots, x_t$ such that each of them is chosen. 
    Notice that this is possible since $t \leq p$. If $t < p$, some
    may be co-located at one $x_i$, though. Then, let $d_1, \ldots, d_q$ be any points on the surface of those balls (i.e., on the spheres).
	Since $t \leq q$ we have enough points to hit all the balls.
	If necessary we can co-locate some points $d_j$ to hit the target
	number $q$. See \figref{cluster}.
\begin{figure}[t]
\centering
\captionsetup{justification=centering}
    \includegraphics[width=0.85\textwidth]{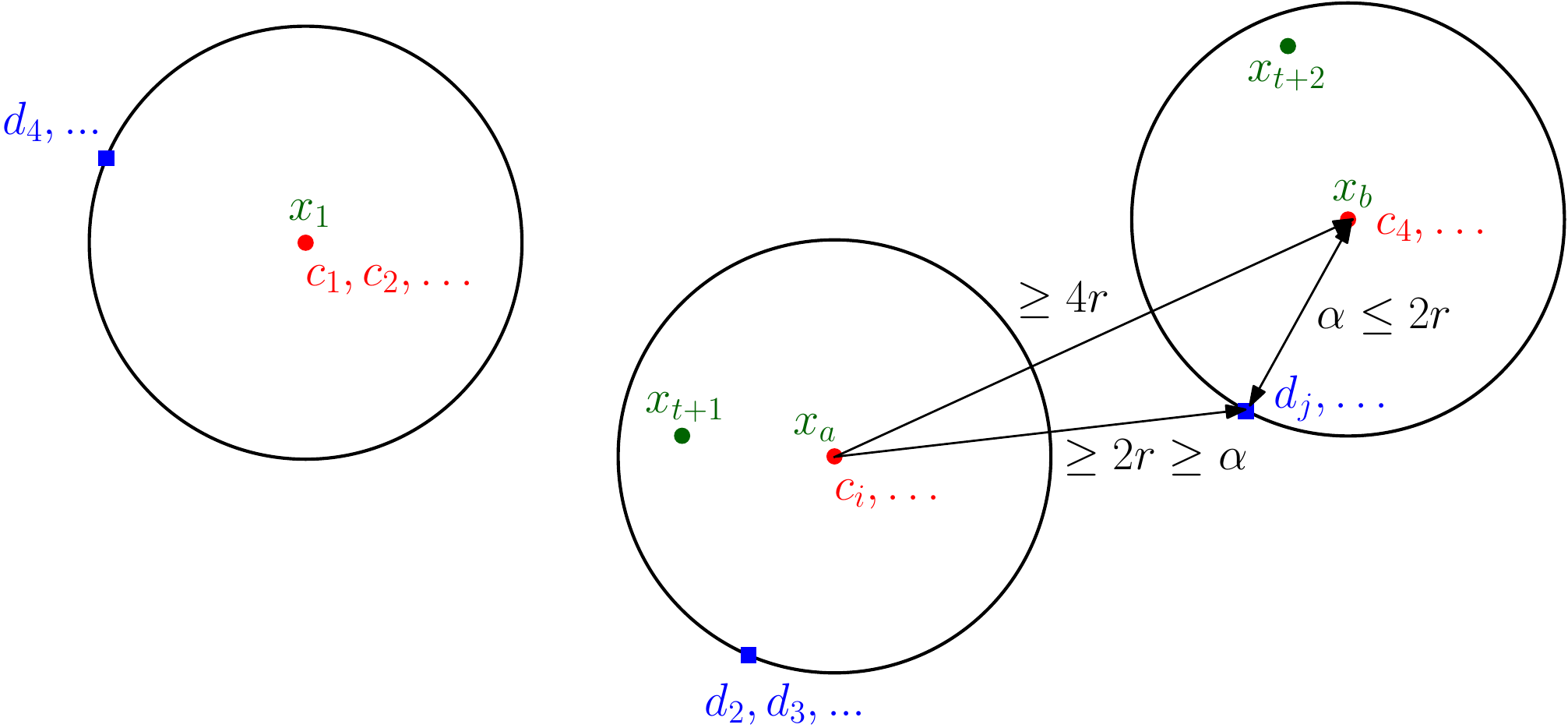}
    \caption{Illustration for proof of \lemref{lem1}}
    \figlab{cluster}%
\end{figure}

    The covering constraints are met since as remarked above,
	the balls of radius $5r$ around all $c_i$ (i.e., around all $x_i$)
	covers $\PntSet$. Similarly, by the triangle inequality and
	because $\alpha \leq 2r$, the balls of radius $7r$ around the $d_j$ cover $\PntSet$.

	To see the separation
	constraint, let $c_i, d_j$ be any red and blue centers 
	as defined above. Suppose $c_i$ is located at the center $x_a$ where $1 \leq a \leq t$, and $d_j$ is located on the surface of the ball $\ball{x_b, \alpha}$ where $1 \leq b \leq t$. If $a = b$, then since $c_i$ is at the
	center and $d_j$ on the surface of the ball $\ball{x_a,\alpha}$ their distance is exactly $\alpha$. If, on the other hand, $a \neq b$, then by the triangle inequality we have that,
	$\dist{x_a, d_j} + \dist{d_j, x_b} \geq \dist{x_a, x_b} \geq 4r$. Now,
	$c_i = x_a$, and $\dist{d_j, x_b} = \alpha \leq 2r$, and so,
	$\dist{c_i, d_j} \geq 4r - 2r = 2r \geq \alpha$, as desired.
\end{proof}
\begin{lemma}
	We have that, $r_{p \land q,\alpha}(\PntSet) \geq \alpha/2$.
\end{lemma}
\begin{proof}
	Consider any point $\pnt_l$. This point is
	in some red ball $\ball{c_i, r}$ and in some
	blue ball $\ball{d_j, r}$. Thus, by the triangle inequality, $\dist{c_i, d_j} \leq \dist{c_i, \pnt_l} + \dist{\pnt_l, d_j} 
	\leq 2r$. On the other hand, $\alpha \leq \dist{c_i, d_j}$. Thus, $\alpha \leq 2r$ and
	the claim follows.
\end{proof}
Observe that since $p \leq q$, $r_p(\PntSet) \geq r_q(\PntSet)$.
\begin{lemma}
	We have that, $r_{p \land q,\alpha}(\PntSet) \geq r_p(\PntSet)$.
\end{lemma}
\begin{proof}
    Consider a feasible solution with
    the radius $\ropt = r_{p \land q, \alpha}(\PntSet)$.
	The $p$ red balls cover $\PntSet$ with
	radius $\ropt$. Thus, $\ropt \geq r_p(\PntSet)$, since by definition,
	$r_p(\PntSet)$ is the minimum $p$-center
	clustering radius.
\end{proof}
Now, for the $O(1)$ approximation to $r_{p \land q,\alpha}(\PntSet)$ notice that
we can easily compute by adapting Gonzalez's algorithm \cite{gonzalez} a $2$-approximation to the $p$-center clustering problem, i.e., to $r_p(\PntSet)$. (This is standard and well-known so we omit the details.)
In other words we have now computed,
$p$ centers $x_1, \ldots, x_p$ all in $\PntSet$, and a radius $r \leq 2r_p(\PntSet)$ such
that balls $\ball{x_i, r}$ cover $\PntSet$. We now consider the
radius $r' = max(r, \alpha/2)$, and clearly
the balls $\ball{x_i, r'}$ also define
such a covering. Now, using \lemref{lem1} we
can find a feasible solution with radius at
most $7r' \leq 14r$. Thus, we have shown the following 
theorem.
\begin{theorem}
   \thmlab{approx-main}%
   Let $\ropt$ be the optimal radius
   for the $\alpha$-separated red blue clustering
   problem on an $n$ point set $\PntSet$ with
   parameters $p, q \geq 1$. Then, we can compute
   in polynomial time, a feasible solution
   where the covering radius is at most
   $14\ropt$.
\end{theorem}
This approximation factor can be improved for the constrained problem, i.e., where all the centers are constrained to be on a fixed line $\Line$. Since this is not our main result, and since we present a polynomial time algorithm for the constrained problem in the next section, we present this in \apndxref{constrained-approx}.

\section{Polynomial time algorithm for the constrained problem}
\seclab{constrained-dp}
Our basic approach will be to do a binary search for the optimal radius. We first present an algorithm to decide
if a given value of the radius $r$ is feasible. Then, we
present an algorithm to determine a finite set of values
such that the optimal radius must be within that set. Then,
a binary search using the feasibility testing algorithm
gives us the optimal radius.
\subsection{Deciding feasibility for given
radius \texorpdfstring{$r$}{r}}
Given a radius $r > 0$ we give a polynomial time method
to decide if there is a solution to the constrained $(n, p \land q, \alpha)$
problem with covering radius $r$. The algorithm is a 
dynamic programming algorithm. One of the challenges 
encountered is that the centers can be anywhere on the line,
and thus a naive implementation of dynamic programming 
does not work since there are not finitely many sub-problems.
As such, we first show how we can compute a finite set
$\CandSet(r)$ of $O(n^2)$ points such that if $r$ is feasible, one can find a
feasible solution with covering radius $r$ with centers
belonging to $\CandSet(r)$.
\subsubsection{Candidate points for the centers}
Consider all the end points of the intervals in $\IntSet(r)$, i.e., $\lept_i(r), \rept_i(r)$, and consider the sorted order of them. Assume that in sorted order the points are 
renamed to $x_1, x_2, \ldots, x_{2n}$ (possibly some of them are co-located). As remarked before, the feasibility problem is equivalent to finding two hitting sets for
the intervals in $\IntSet(r)$ that satisfy the separation
constraints. As in standard in such hitting set problems, we look at the \emph{faces} of the arrangement of the intervals. Some of these faces might be open 
intervals, or half open intervals, or even singleton points. Notice that all faces are disjoint by definition. It is easy to see that if $F$ is 
such a face, then a point in the closure $\cl{F}$ of $F$ will hit
at least the same intervals as points in $F$ hit. In order to avoid confusion when we refer to faces vs. their closures, in the remaining discussion we will always say face $F$ for the original face and closure face $\cl{F}$ when referring to one of the closures (even though they may be the same set of points).

We explain now, why considering face closures is valid. Suppose a certain center belongs to a
face $F$ and suppose it is allowable to choose \emphi{any} point close enough to
one of its boundary points, such that the separation constraints are met. Then, it is also valid to choose
it at the boundary point and respecting the separation constraint since the
separation constraint is that distance between the red and blue centers is
$\geq \alpha$ as opposed to a strict inequality $> \alpha$. Therefore, it is valid to replace faces by their closures.

To compute all the closures of the faces in the arrangement 
of the $\IntSet(r)$, we sort the $x_i$ and we retain all the
consecutive intervals $[x_i, x_{i+1}]$ that do not lie outside any of the intervals $\Int_i(r)$. This
can be done by a simple line sweep algorithm. Notice that
there are only $O(n)$ such closure faces.

Next, we define 
a sequence for each such closure face. Consider such a closure face, $[x_i, x_{i+1}]$ that is the \emphi{starting closure face} for this sequence. Consider the sequence, $x_i, x_i + \alpha, x_i + 2\alpha, \ldots$. We only want to retain, for each closure face,
(at most) the first three points that hit the closure face. So, 
given any starting closure face $[x_i, x_{i+1}]$ there are only $O(n)$ points in this sequence since it is bounded by the number of closure faces (in fact beyond the starting closure face $[x_i, x_{i+1}]$) times three. Let the sequence of points that result due to starting closure face
$[x_j, x_{j+1}]$ be denoted by $\seq_j$.
We have the following lemma,
\begin{lemma}
\lemlab{seq1}%
	For each starting closure face, $[x_i, x_{i+1}]$ the associated
	sequence $\seq_j$ can be computed in $O(n)$ time.
\end{lemma}

\begin{proof}
We consider each closure face
and compute the points of this sequence that possibly lie
in this closure face. Consider such a closure face $[x_j, x_{j+1}]$. For
a member of $\seq_i$ to lie in this closure face there is an
integer $k$ such that $x_j \leq x_i + k\alpha \leq x_{j+1}$.
This is equivalent to,
$\frac{x_j - x_i}{\alpha} \leq k \leq \frac{x_{j+1} - x_i}{\alpha}$. Thus, to find the first three points of $\seq_i$ hitting the closure face, we only need
to find the three smallest integers in the interval,
$[\frac{x_j - x_i}{\alpha}, \frac{x_{j+1} - x_i}{\alpha}]$,
if there are such. This can be done in $O(1)$ time per closure face. Since there are $O(n)$ closure faces, the entire
sequence can be constructed in $O(n)$ time.
\end{proof}

Computing such sequence for each closure face as starting closure face leads to a total of $O(n^2)$ points, and can be computed in $O(n^2)$ time overall by the previous lemma. The set $\CandSet(r)$ is the set of the points in all these sequences. Let the sorted order of points in this set be denoted by
$\cpt_1, \cpt_2, \ldots, \cpt_m$ where $m = O(n^2)$. Thus $\CandSet(r) = \{ \cpt_1, \ldots, \cpt_m\}$.
For any such point $\cpt_k$ denote by $\nxt{k}$ the first index $j$ such that $\cpt_j - \cpt_k \geq \alpha$. If there is no such point, let $\nxt{k} = m + 1$. Notice that we 
can compute $\nxt{k}$ by a successor query in $O(\log n)$
time if the set $\CandSet(r)$ is sorted. 
The following lemma says that we can assume that the 
centers (of both colors) are in $\CandSet(r)$.
\begin{lemma}
	Suppose that the constrained $(n, p \land q, \alpha)$ problem has a feasible solution with covering radius $r$.
	Then there is a solution with all centers
	belonging to $\CandSet(r)$.
\end{lemma}

\begin{proof}
	Consider a feasible solution with $p$ red centers and $q$ blue 
	centers. First, we remark that we can assume, wlog, that in any face $F$ there are at most two points, one red and one blue. This is true
	because having more than one red or more than one blue point in a face does not affect the covering constraints, as each point in a face hits (i.e., belongs to) the exact same set of intervals by definition. Thus, we may assume that there are at most $T \leq (p + q)$ such centers, since we might need to throw away some of them when two centers of the same color belong to one face. Let the $T$ centers be $u_1, \ldots, u_{T}$ where any of them can be red or blue. We show how to construct iteratively another feasible
	solution where all the centers are in $\CandSet(r)$.

	We will proceed face by face, and consider all the centers within 
	the face. We know that there are at most two centers within a face.
	Moreover, if there are two they are of different color.
	Our basic strategy is to move the first center left till we can,
	while remaining within the closure of the face, without violating
	any separation constraint. If there is only one point in a face,
	we are done. Otherwise, once the position of the first point is
	fixed, the second point can be similarly moved left until its position is determined within $\CandSet(r)$. Let the sorted
	order of faces be $F_1, F_2, \ldots, F_N$ where $N = O(n)$. 
	
	We construct a new sequence $v_1, \ldots, v_T$ where $v_i$ is
	assigned color of $u_i$ and is obtained by shifting $u_i$ to the
	left (so that it will lie in $\CandSet(r)$, but never leaving
	closure of the face it belongs to. We prove the following
	claim by induction, which implies the claim that all the $v_i$
	belong to $\CandSet(r)$.
	
	\begin{claim}
	    For each $k \geq 1$, all the points $u_i$ belonging
	    to face $F_k$ are mapped to points $v_i$ (belonging to
	    $\cl{F}_k$) such that, the at most two such $v_i$, lie on
	    consecutive points of the same sequence $\seq_j$ for some
	    $j$.
	\end{claim}

    Consider the base case $k = 1$. If there are no points in $F_1$,
    the claim holds vacuously. If there is only one point in $F_1$,
    slide it left until it hits the boundary of $F_1$. This does not violate any constraints. The claim holds
    true trivially. Suppose there are two points in $F_1$. Now,
    $\cl{F}_1 = [x_1, x_2]$ and after sliding the first point left
	till it coincides with $x_1$, and thus in $\seq_1$, the second
	point clearly satisfies $u_2 - x_1 \geq \alpha$ since even before
	the sliding the inequality was satisfied. Notice that the points
	are of different color. Thus we can place the second point $v_2$ 
	at $x_1 + \alpha \in \seq_1$ and they are consecutive points of $\seq_1$.
	
	Suppose that the claim is true up-to $k$. We now consider
	the case $k+1$. Again, as before if there are no points in 
	$F_{k+1}$ the claim holds vacuously. If there is only 
	one point, we slide it left to the first point in $\CandSet(r)$
	which is allowable for it. The meaning of allowable is the following. Suppose this point is $v_s$. Then, if $v_{s-1}$, which
	is in a previous face, is of the same color as $v_s$, then $v_s$
	can be anywhere within its face. If it is of different color, 
	then $v_s$ has to be at least at $v_{s-1} + \alpha$. Since
	the starting point of the closure $\cl{F}_{k+1}$ is in
	$\CandSet(r)$, as is $v_{s-1} + \alpha$ if it lies in $\cl{F}_{k+1}$, while sliding left we will hit a point in $\CandSet(r)$ eventually and we stop there. Now consider,
	the case where there are two points $u_i, u_{i+1}$ in $F_{k+1}$. After $v_i$ has been placed at its position in $\CandSet(r)$
	as outlined for the case of single point in $F_{k+1}$,
	suppose it belongs to $\seq_j$ for some $j$. Clearly it is at 
	most the second point, of $\seq_j$ in $\cl{F}_{k+1}$ as the second point is already $\alpha$ ahead of the first point of $\seq_j$ in $\cl{F}_{k+1}$. Now, $u_{i+1} - v_i \geq \alpha$. Thus, we can place the second point of
	$F_{k+1}$ at the next point of $\seq_j$ in $\cl{F}_{k+1}$, which exists in $\cl{F}_{k+1}$ since the next point is $v_i + \alpha$ 
	which is in $\cl{F}_{k+1}$ by assumption. We observe that all claims hold
	true.
\end{proof}

\subsubsection{The dynamic programming algorithm}
The dynamic programming algorithm computes two tables,
$\ccr[\IntSet_1,\IntSet_2,a,b,k]$, 
and $\ccb[\IntSet_1,\IntSet_2,a,b,k]$ of Boolean $\true, \false$. Here, $\IntSet_1, \IntSet_2$ are prefixes of intervals of $\CandSet(r)$ (when they are ordered by their left and right end-points), and $0 \leq a \leq p, 0 \leq b \leq q$ are integers,
and $1 \leq k \leq (m + 1)$ is also an integer. The table entry 
$\ccr[\IntSet_1, \IntSet_2, a, b, k]$ is
$\true$, if there is a hitting set consisting of (at most) $a$ red points that hit
the intervals in $\IntSet_1$, (at most) $b$ blue points that hit the intervals
in $\IntSet_2$ and with the constraint that the first point to be possibly put is red and at $\cpt_k$ if $k < m$. Here $k > m$ represents that there is no where to really put the 
first red point. Notice that the separation constraint between red/blue
points must be met. Similarly the table entry $\ccb[\IntSet_1, \IntSet_2, a, b, k]$ is true if the first point is blue and at 
$\cpt_k$ (for $k \leq m$). Assuming that the above tables have been
computed, we can answer whether the radius $r$ is feasible by
computing the following expression, where we denote $\IntSet(r)$ by
$\IntSet$ for brevity,
\begin{align*}
\ccr[\IntSet,\IntSet,p,q,1] \lor & \ldots \lor \ccr[\IntSet,\IntSet,p,q,m]\\
&\bigvee\\
\ccb[\IntSet,\IntSet,p,q,1] \lor & \ldots \lor \ccb[\IntSet,\IntSet,p,q,m].
\end{align*}
In the above expression we try to hit all the intervals in $\IntSet$
by both red and blue points and we try all possible starting locations
and color for the first point. We know that the centers can be assumed
to belong to $\CandSet(r) = \{\cpt_1, \ldots, \cpt_m\}$.

Now we present the recursive definition of the algorithm to fill the tables.
We only present the definitions for $\ccr[\cdot]$ but there
is an entirely similar definition for $\ccb[\cdot]$ with
the roles of red and blue interchanged.
\begin{gather*}
\ccr[\IntSet_1,\IntSet_2,a,b,k] = \\
\begin{cases}
\false & \text{ if } (\IntSet_1 \neq \emptyset \land  a = 0) \lor (\IntSet_2 \neq \emptyset \land b = 0) \lor (\IntSet_1 \cup \IntSet_2 \neq \emptyset \land k > m), \\
\true & \text{ if } (\IntSet_1 = \emptyset \land \IntSet_2 = \emptyset), \\
\false & \text{ if there exists an interval in } \IntSet_1 \cup \IntSet_2 \text{ ending before } \cpt_k,\\
B_a \lor B_b & \text{ otherwise}.
\end{cases}
\end{gather*}
The first case means that if there are not any red centers
to put while some unhit intervals remain in $\IntSet_1$, or 
not any blue centers to put but unhit intervals in $\IntSet_2$, or if we have already passed over all
centers ($k > m$) but any unhit red or blue intervals remain,
we return false. The next case means that if that all intervals have been hit already we should return true. The
penultimate case means that if the first point $\cpt_k$ is so
far ahead that at least one interval in $\IntSet_1 \cup \IntSet_2$ ends before it, then there can be no solution.
This is true because any later points, red or blue, will
only be ahead of $\cpt_k$ and thus the ended interval cannot
be hit. The last case uses Boolean variables $B_a, B_b$ that
are defined as follows, and they also capture the
main recursive cases. As required by the definition
of the function, we must put the next center as red
and at $\cpt_k$. This would cause some intervals
in $\IntSet_1$ to be hit by $\cpt_k$. We remove those
intervals from $\IntSet_1$. Let $\IntSet'_1$ be the
intervals in $\IntSet_1$ not hit by $\cpt_k$. It is easy to see that if $\IntSet_1$ is a prefix of $\IntSet$, then so is $\IntSet'_1$. The definitions
of $B_a, B_b$ are as follows.
\[
B_a \gets (\IntSet_1 = \emptyset) \lor \bigvee_{j=k+1}^m \ccr[\IntSet'_1,\IntSet_2,a-1,b,j]
\]
This assignment ensures that if $\IntSet_1 = \emptyset$, then
we never really try to put any red point. If not, then
we try all possibilities for the next position of the red 
point. Notice that putting another red point at
$\cpt_k$ is not necessary so we start with the remaining
positions and go up-to $m$. The coverage requirements for blue
points and their numbers remain unchanged. The red number 
decreases by $1$. The Boolean $B_b$ has the following
definition,
\[
B_b \gets (\IntSet_2 = \emptyset) \lor 
\bigvee_{j = \nxt{k}}^m \ccb[\IntSet'_1,\IntSet_2,a-1,b,j].
\]
This is because, if the next point (after the current red one) is to be blue, it can only be at index $\nxt{k}$ or later. Thus we look-up $\ccb[\IntSet'_1, \IntSet_2,a-1,b,j]$ for
all such possible $j$. The first check $\IntSet_2 = \emptyset$ means that if the blue intervals have already been
hit, we do not need to put any blue point later. Both the
tables $\ccr[\cdot], \ccb[\cdot]$ are filled simultaneously by first filling in the entries
fitting the base cases, and then traversing them in order
of increasing $a$, increasing $b$, decreasing $k$, and
increasing $\IntSet_1, \IntSet_2$ (i.e., the smaller prefixes
come earlier). It is easy to see that the traversal order
meets the dependencies as written in the recursive definitions.

\noindent \textbf{Analysis.} 
First observe that computing the candidate centers
can be done in $O(n^2)$ time as implied by \lemref{seq1}
and the following discussion. Moreover, the successor
points $\nxt{k}$ can all be computed in total 
$O(n^2 \log n)$ time by first sorting $\CandSet(r)$
and then followed by successor queries. The time
however is dominated by the main dynamic programming
algorithm.
Observe that there
are $O(n)$ prefixes, and $m = O(n^2)$ possible center
locations. Thus there are in total $O(n^4 p q)$ entries 
to be filled. Except for the base cases, filling in an 
entry requires looking up $O(n^2)$ previous entries,
as well as some computation such as finding which intervals
are not hit by the current point. Such queries can be handled
easily for all the intervals say in $\IntSet_1$ wrt the
point $\cpt_k$ in $O(n)$ time. Thus for a particular
table entry we require $O(n^2)$ time. Overall we will 
take $O(n^6 p q)$ time. We get the following theorem,
\begin{theorem}
\thmlab{feasibility-main}%
For the constrained $(n, p \land q, \alpha)$ problem where
the centers are constrained to lie on the $x$-axis, given
a radius $r$, it can be decided if $r$ is feasible in time
$T_{DP}(n, p, q) = O(n^6 p q)$. Moreover, if $r$ is feasible,
a feasible solution with covering radius $r$ can also be
computed in the same time.
\end{theorem}
To justify the comment about the feasible solution, note 
that by standard dynamic programming techniques, we can
also remember while computing the table entries the solution,
and it can be output at the end.
\subsection{Candidate values for \texorpdfstring{$r$}{r}}
\seclab{rcandidates}%
In this section, we will find a discrete candidate set for the optimal radii  that facilitates a polynomial time algorithm for solving the constrained $(n, p\land q, \alpha)$ problem as presented in \secref{main-result}. For this purpose, we need to determine some properties of an optimal solution.
First, we define a standard form solution and describe an easy approach to convert a feasible solution to standard form. Then we present a lemma for proving a property of an optimal solution. Finally, we compute a finite candidate set for optimal radii.

Let $U=\{u_1, u_2, \dots, u_{p+q}\}$ be a feasible solution with covering radius $r$. The closure face that contains $u_i$ is denoted by $[x_{i,1}(r),x_{i,2}(r)]$, where $x_{i,1}(r)$ and $x_{i,2}(r)$ are the endpoints of some intervals (i.e., $a_j(r)$ or $b_j(r)$).  Since $u_i$s are on the $x$-axis, be a slight abuse of notation, we let $u_i$ denotes the $x$-coordinate of point $u_i$.
For a given feasible solution $U=\{u_1, u_2, \dots, u_{p+q}\}$, its \emphi{standard} form has two following properties:

1. If $u_1$ and $u_{p+q}$ are on the endpoints.

2. Any two consecutive same color centers are on the endpoints.

\noindent {\bf Converting a given solution to standard form:}
If $u_1$ (resp. $u_{p+q}$) is not on an endpoint, we move it to the left (resp. right) to hit  $x_{1,1}(r)$ (resp. $x_{p+q,2}(r)$). For every pair of two consecutive same color  centers $u_i$ and $u_{i+1}$, $1\leq i\leq p+q-1$, if $u_i$ is not on an endpoint, we move it to the right to hit $x_{i,2}(r)$ and if $u_{i+1}$ is not on an endpoint we move it to the left to hit $x_{i+1,1}(r)$.

Clearly, the standard form solution as constructed above satisfies the covering and separation constraints.
Let $S(k,j)$ denote a sequences of $j+1$ consecutive centers in $U$ starting from $u_k$, i.e.,  $(u_k, u_{k+1},\dots , u_{k+j})$. A sequence $S(k,j)$ is called \emphi{alternate} if for all $i, k\leq i\leq k+j-1$, $u_i$ and $u_{i+1}$ have different colors and centers $u_k$ and $u_{k+j}$ are on the endpoints and the other centers of the sequence are not on the endpoints (such a center is called \emphi{internal}).

Now note that if $U$ is a standard solution, then the consecutive red-blue centers can be clustered in some alternate sequences.
These alternate sequences can be provided by scanning the centers from left to right and clustering a couple of consecutive blue-red centers between two endpoints that include a center. To this end, we have the following simple approach:

\noindent {\bf Clustering centers in alternate sequences:}
Let $u_{cur}$ be the first center that has not been visited yet. At the beginning, $u_{cur}=u_1$.  Let $u_i$ be the next closest different color center to $u_{cur}$.
 All centers from $u_{cur}$ to $u_{i-1}$ should be on the endpoints since they are the same color.
 We can construct the next sequence from $k=i-1$, i.e., we add $u_{i-1}$ and $u_i$ to a sequence.  There are two events:

 {\bf Event 1}:  $u_i$ is on an endpoint, so the sequence is completed. If there are any unvisited centers, we continue scanning the centers by starting  from $u_i$, i.e., we mark $u_i$ as unvisited, set $u_{cur}=u_i$ and proceed as before until there is no unvisited center.

{\bf Event 2}:  $u_i$ is not on an endpoint. Consider $u_{i+1}$. 
$u_{i+1}$ and $u_i$ should have different colors since $U$ is standard.  We add $u_{i+1}$ to the sequence. If there are any unvisited centers, consider $u_{i+2}$ and check Events 1 or 2 for $i=i+2$.

Note that some of the centers may belong to two alternate sequences (e.g., a center on an endpoint with different color adjacent centers) and some of them may not be in a sequence (e.g., a center with  same-color adjacent centers).

Now we prove a property of the optimal solutions in standard form for being able to find a discrete candidate set for the optimal radii.
\begin{lemma}
\lemlab{candidate}
 Let $U=\{u_1, u_2, \dots, u_{p+q}\}$ be a feasible solution for the constrained $(n,p\land q,\alpha)$ problem with  radius of covering $r$. If the distance between any two endpoints of the intervals in $\CandSet(r)$ is not
 $t \alpha$, where $t\in \mathbb{Z}, 0\leq t\leq p+q-1$, then the constrained $(n,p\land q,\alpha)$ problem has a feasible solution with radius less than  $r$.
\end{lemma}
\begin{proof}
We will show that there is a real number $0<\epsilon<r$ such that the constrained $(n,p\land q,\alpha)$ problem has a feasible solution with radius of covering  $r-\epsilon$. To this end, we obtain a set of centers, $\bar{U}$, from the given feasible solution $U$ and show that the set of balls centered at the points in  $\bar{U}$ with  radius $r-\epsilon$ is a feasible solution for the problem. First, we need to modify $U$ to find a feasible solution with the property that any two consecutive blue and red centers are at a distance  strictly greater than $\alpha$ (not exactly $\alpha$). Then we use it for finding a solution, $\bar{U}$, with radius of covering $r-\epsilon$ (that is explained later).

First of all, we convert $U$ to standard form and compute all alternate sequences of the standard solution. Then we use them to find a feasible solution with the property that any two consecutive blue and red centers are at a distance of strictly greater than $\alpha$. Note that by the Lemma's assumption, each alternate sequence  $S(k,j)$ has at least a pair of two consecutive centers at a distance of strictly greater than $\alpha$ (since each distance is at least $\alpha$ and sum of them is not $j\alpha$). But we need to have this strict inequality for all such pairs. So in each sequence $S(k,j)$, if there are two consecutive centers $u_{k+i}$ and $u_{k+i+1}$ an a distance of exactly $\alpha$, we should perturb the internal centers such that the distance between any two consecutive blue and red centers is strictly greater than $\alpha$.

For perturbing the internal centers of each alternate sequence $S(k,j)$, if $j=1$, then $u_{k+1}-u_k >\alpha$, because $u_k$ and $u_{k+1}$ are on the endpoints so $u_{k+1}-u_k \neq \alpha$. If $j>1$, we proceed by induction on the number of the pairs with the distance of $\alpha$ which is denoted by $n_{k,j}$. For $n_{k,j}=1$,  let $u_{k+i}, u_{k+i+1} \in S(k,j)$ such that $u_{k+i+1}-u_{k+i} =\alpha$. Since $j>1$, at least one of $u_{k+i}$ and $u_{k+i+1}$ is internal, say $u_{k+i}$. Since $n_{k,j}=1$, $u_{k+i}-u_{k+i-1} >\alpha$. So we can shift $u_{k+i}$ toward $u_{k+i-1}$ infinitesimally such that $u_{k+i}-u_{k+i-1} >\alpha$  and we still have $u_{k+i+1}-u_{k+i} >\alpha$.  Assume for the induction hypothesis that for all integers $m>1$, in a sequence $S(k,j)$ with $n_{k,j}<m$, including a pair of consecutive red and blue centers at a distance greater than $\alpha$, we can perturb the internal centers such that all distances between two consecutive centers are strictly greater than $\alpha$. Now assume that $n_{k,j}=m>1$. For some $0\leq i\leq j-1$, let $u_{k+i}, u_{k+i+1} \in S(k,j)$ such that $u_{k+i+1}-u_{k+i}=\alpha$.  $S(k,j)$ has a pair of two consecutive centers at a distance greater than $\alpha$. This pair belongs to one of the sequences $S(k,i)$ or $S(k+i+1,j)$, say $S(k,i)$. It is clear that $n_{k,i} <m$, so by the induction hypothesis, we can perturb the internal centers of $S(k,i)$ such that all distances between two consecutive centers are strictly greater than $\alpha$. Next, we can move $u_{k+i}$ toward $u_{k+i-1}$ infinitesimally such that $u_{k+i}-u_{k+i-1} >\alpha$  and we now also have $u_{k+i+1}-u_{k+i}>\alpha$.  Now we add $u_{k+i}$ to $S(k+i+1,j)$ to obtain $S(k+i,j)$ in which the distance between centers $u_{k+i}$ and $u_{k+i+1}$ is greater than $\alpha$. Since $n_{k+i,j} <m$, again by the induction hypothesis, we can perturb the internal centers of $S(k+i,j)$ such that all distances between two consecutive centers are strictly greater than $\alpha$. It means that $S(k,j)$ no longer contains a consecutive pair with distance $\alpha$.

Now we can compute $\bar{U}$. Let $0<\epsilon<r$ be a positive 
real number to be fixed later. If $u_i$ is on an endpoint, say
$x_{i,1}(r)$, let $\bar{u_i}= x_{i,1}(r-\epsilon)$, otherwise,
$\bar{u_i}=u_i$. Notice that by our assumptions, there is no solution
for $t = 0$ so all endpoints are distinct. As such for an $u_i$ on an endpoint, it
is never on two endpoints simultaneously and its movement is unambiguously
determined.
We will show that there exists an $\epsilon$ such 
that $\bar{U}=\{\bar{u}_1,\bar{u}_2, \dots , \bar{u}_{p+q}\}$ is a
feasible solution with  radius of covering  $r-\epsilon$, i.e.,
$\bar{U}$ should satisfy the covering and separation constraints. To
this end, firstly, after decreasing $r$ to $r-\epsilon$, the relative
order of the endpoints of the intervals should not change, i.e., the
displacement of an endpoint of a face $F_i$ should be less than
$||F_i||/2$, where $||F_i||$ is the distance between the endpoints of
face $F_i$. Secondly, for satisfying the covering constraint, the
internal centers should remain in their faces, i.e.,
$x_{i,1}(r-\epsilon)<\bar{u}_i<x_{i,2}(r-\epsilon)$. So the
displacement of point  $x_{i,1}$  (resp. $x_{i,2}$) should be less 
than $u_i-x_{i,1}(r)$ (resp. $x_{i,2}(r)-u_i$). Finally, for satisfying
the separation constraint, in each sequence $S(k,j)$, we should have
$\bar{u}_{k+1}-\bar{u}_k\geq \alpha$ and
$\bar{u}_{k+j}-\bar{u}_{k+j-1}\geq \alpha$ (note that the distance
between two internal centers does not change), i.e., the displacement
of the endpoint that contains $u_k$ (resp. $u_{k+j}$) should be less
than $u_{k+1}-u_k-\alpha$ (resp. $u_{k+j}-u_{k+j-1}-\alpha$).
Therefore, by choosing real numbers $\delta_1$, $\delta_2$, $\delta _3$ as follows, and
$0<\delta <\min \{\delta_1, \delta_2, \delta_3\}$,
because of continuity of the movement of endpoints on line  $\Line$, we can obtain a positive $\epsilon$ such that the displacement of an endpoint becomes at most $\delta$ when the radius decreases to $r-\epsilon$.
\begin{gather*}
0<\delta _1< 1/2 \min_{1\leq i\leq 2n-1}\{||F_i||\} \\
0<\delta_2<\min_{\forall S(k,j)} \{ \min_ {k+1\leq i\leq k+j-1} \{ u_i-x_{i,1}(r), x_{i,2}(r)-u_i \}\} \\
0<\delta_3<\min_{\forall S(k,j)} \{ u_{k+1}-u_k-\alpha , u_{k+j}-u_{k+j-1}-\alpha \}
\end{gather*}
Consequently, there exists a non zero $\epsilon>0$ such that the balls centered at points in $\bar{U}$ with covering radius $r-\epsilon$ is a feasible solution.
\end{proof}
By \lemref{candidate}, in the optimal solution, there is at least a pair of two endpoints at distance $t\alpha$, where $t\in \mathbb{Z}, 0\leq t\leq p+q-1$.
The interval
endpoints $\lept_i(r)$ and $\rept_i(r)$ are given by $\lept_i(r) = p_{i1} - \sqrt{r^2 - \sum_{j=2}^d p_{ij}^2}$,
and $\rept_i(r) = p_{i1} + \sqrt{r^2 - \sum_{j=2}^d p_{ij}^2}$, so, a candidate set for the optimal radius can be computed by solving the following equations for all $1\leq i,k \leq n$ and $t\in \mathbb{Z}, 0\leq t\leq p+q-1$:
$$ \pnt_{i1} \pm \sqrt{r^2 - \sum_{j=2}^d p_{ij}^2}-\pnt_{k1} \pm \sqrt{r^2 - \sum_{j=2}^d p_{kj}^2}=t \alpha,$$ since at least one of those equalities holds true. 
Due to our general position assumption, i.e., no two points in $\PntSet$ have the same distance from $\ell$, these equations have a finite number of solutions. 
This is not too hard to show, but for completeness we show the details in \apndxref{eqsolve}. The general position assumption can be removed. Without the
assumption, \lemref{candidate} needs an amended statement and proof. Due to space
constraints, we show the details in \apndxref{nogp}.

By solving these equations, we obtain $O(n^2(p+q))$ candidates for the optimal radius and this proves the following lemma.

\begin{lemma}
\lemlab{finite-sol}%
There is a set of $O(n^2(p + q))$ numbers, such that
the optimal radius $r_{p\land q, \alpha}(\PntSet)$ is
one among them, and this set can be constructed in
$O(n^2(p+q))$ time.
\lemlab{ncandidates}
\end{lemma}
\subsection{Main result}
\seclab{main-result}%
By first computing the candidates for $r^*$ and then performing a binary search over them using the feasibility testing algorithm, we can compute the optimal radius. Thus we have the following theorem.
\begin{theorem}
\thmlab{constrained-main}%
The constrained $(n,p\land q, \alpha)$ problem can be solved in $O(n^2(p+q) + T_{DP}(n,p,q)\log n) = O(n^6 p q \log n)$ time.
\end{theorem}

\section{Conclusions}
Improving the approximation factor of our main 
approximation algorithm (\thmref{approx-main}) and the running time of our polynomial time algorithm for the constrained 
problem (\thmref{constrained-main}) are obvious candidates for problems for future research work. Apart from this it seems that a multi-color generalization of the $k$-center problem
is worth studying for modeling similar practical
applications. Here we want $k$ different colored
centers, and balls of each color covering all of $\PntSet$
but with the separation constraints more general, i.e.,
between the centers of colors $i, j$ the distance must
be at least some given $\alpha_{ij}$. It seems that new
techniques would be required for this general problem.
\seclab{conclusions}
\newpage
\bibliography{bibliography}

\begin{thebibliography}{10}

\bibitem{agarwal}
P.~K. Agarwal and C.~M. Procopiuc.
\newblock Exact and approximation algorithms for clustering.
\newblock {\em Algorithmica}, 33(2):201--226, 2002.

\bibitem{boselangerman}
P.~Bose, S.~Langerman, and S.~Roy.
\newblock Smallest enclosing circle centered on a query line segment.
\newblock In {\em Proc. of Can. Conf. on Comp. Geom.}, 2008.

\bibitem{bosetoussaint}
P.~Bose and G.~Toussaint.
\newblock Computing the constrained euclidean geodesic and link center of a
  simple polygon with applications.
\newblock In {\em Proc. Pacific Graph. Int.}, pages 102--112, 1996.

\bibitem{brass}
P.~Brass, C.~Knauer, H.-Suk Na, C.-Su Shin, and A.~Vigneron.
\newblock The aligned $k$-center problem.
\newblock {\em Int. J. Comp. Geom. Appl.}, 21(2):157--178, 2011.

\bibitem{das}
G.~Das, S.~Roy, S.~Das, and S.~Nandy.
\newblock Variations of base-station placement problem on the boundary of a
  convex region.
\newblock {\em Int. J. Found. Comput. Sci.}, 19:405--427, 2008.

\bibitem{gonzalez}
T.~F. Gonzalez.
\newblock Clustering to minimize the maximum intercluster distance.
\newblock {\em Th. Comp. Sc.}, 38:293--306, 1985.

\bibitem{hurtado}
F.~Hurtado and G.~Toussaint.
\newblock Constrained facility location.
\newblock {\em Studies of Location Analysis, Sp. Iss. on Comp. Geom.}, pages
  15--17, 2000.

\bibitem{hwang}
R.~Z. Hwang, R.~C.~T. Lee, and R.~C. Chang.
\newblock The slab dividing approach to solve the euclidean $p$-center problem.
\newblock {\em Algorithmica}, 9:1--22, 1993.

\bibitem{kavand}
P.~Kavand, A.~Mohades, and M.~Eskandari.
\newblock $(n,1,1,\alpha)$-center problem.
\newblock {\em Amirkabir Int. J. of Sc. \& Res.}, 2014.

\bibitem{megiddo}
N.~Megiddo.
\newblock Linear time algorithms for linear programming in {$\Re^3$}.
\newblock {\em SIAM J. Comput.}, 12(4), 1983.

\bibitem{megiddosupotwik}
N.~Megiddo and K.~J. Supowit.
\newblock On the complexity of some common geometric location problems.
\newblock {\em SIAM J. Comput.}, 13(1):182--196, 1984.

\bibitem{roy}
S.~Roy, D.~Bardhan, and S.~Das.
\newblock Efficient algorithm for placing base stations by avoiding forbidden
  zone.
\newblock In {\em Proc. of the Sec. Int. Conf. Dist. Comp. and Int. Tech.},
  page 105–116, 2005.

\bibitem{shin}
C.-S. Shin, J.-H. Kim, S.~K. Kim, and K.-Y. Chwa.
\newblock Two-center problems for a convex polygon.
\newblock In {\em Proc. of the 6th Ann. Euro. Symp. Alg.}, pages 199--210,
  1998.

\bibitem{sylvester}
J.~J. Sylvester.
\newblock A question in the geometry of situation.
\newblock {\em Quart. J. Math.}, 322(10):79, 1857.

\end{thebibliography}

\newpage
\appendix

\section{Approximation algorithm for the constrained \texorpdfstring{$(n,p \land q, \alpha)$}{npqa} Problem}
\apndxlab{constrained-approx}%
Here we provide an approximation algorithm for the constrained  $(n, p \land q, \alpha)$ problem
improving the result of \thmref{approx-main}. Without loss of generality, we assume that $\Line$ is the $x$-axis. We present a 4-approximation algorithm for the constrained $(n, p\land q, \alpha)$ problem with running time $O(n \log ^2 n)$.

First, we find an optimal solution for the constrained $p$-center problem. We denote the optimal radius by $r^*_p$ and lets the balls of radius $r^*_p$ be $B_1, B_2, \ldots, B_p$ from left to right sorted by the order of $x$-coordinates of their centers on $x$-axis. If $r^*_p<\alpha/2$, we expand  balls $B_i$ by a factor of $\alpha/(2r^*_p)$.
We would like to divide the balls into some clusters denoted by $G_i$. The first cluster, $G_0$, contains all balls that intersect $B_1$. $G_{i+1}$ is the set of all balls that intersect the leftmost ball which is not in $G_i$.

Now for finding a solution for the constrained $(n, p\land  q, \alpha)$ problem, for each cluster, we output two balls, one red and one blue, with a distance of at least $\alpha$
among their centers, to cover the points in the cluster. The algorithm is presented in
\algoref{conalgo1}, and $CORE[i]$ refers to the ball that
defines the cluster $G_i$, i.e., it is the leftmost ball
not in $G_{i-1}$.
\begin{figure}
\centering
\begin{minipage}{.9\textwidth}
 \centering
 \small{
\begin{algorithm}[H]
\DontPrintSemicolon

  \KwInput{A set $\PntSet$ of $n$ points, a horizontal line $\Line$, integers $p$ and $q$, $p\leq q$, and a real number $\alpha \geq 0$.}
  \KwOutput{A set of $p$ balls centered at $c_{1}, c_{2},\ldots, c_{p}$ on $x$-axis that covers $\PntSet$ and a set of $q$ balls centered at $d_{1}, d_{2},\ldots , d_{q}$ on $\Line$ that covers $\PntSet$ such that for all $1\leq i\leq p$ and $1\leq j\leq q$, $|c_i-d_j|\geq  \alpha$.}

  $\{B_1, B_2, \dots, B_p\}$ is an optimal solution for the constrained $p$-center problem with radius $r^*_p$.

  $m \gets 0, j \gets 1, i \gets 1, G_0 \gets \emptyset$

  \While{$j\leq p$}
   {
   $CORE[m] \gets B_j$

   $i \gets i + 1$

  \While {$B_i$ intersects $B_j$}
  {
  $G_m \gets G_m\cup \{B_i\}$

    $i \gets i + 1$
  }
  $j \gets i$

  $m \gets m + 1$

  $G_m \gets \emptyset$ \;
  }

 $i \gets 0$

 \While {$i< m$}
 {
 \If {$i\mod 2 = 0$} {$c_{i+1} \gets $ leftmost point of $CORE[i]$

 $d_{i+1} \gets $ rightmost point of $CORE[i]$ }

\Else {$c_{i+1} \gets$ rightmost point of $CORE[i]$

 $d_{i+1} \gets $ leftmost point of $CORE[i]$
 }
  }
  \If {$r^*_p<\alpha/2$}{\Return  $\{\ball{c_i,2\alpha}, \ball{d_i,2\alpha}\;|\; 1 \leq i\leq m\}$ }
  \Else { \Return  $\{\ball{c_i,4r^*_p}, \ball{d_i,4r^*_p}\; |\; 1\leq i\leq m\}$}
\caption{4-approximation algorithm for constrained $(n, p\land q, \alpha)$-center problem}
\algolab{conalgo1}%
\end{algorithm}
}
\end{minipage}
\end{figure}

\noindent \textbf{Correctness.}
Notice that all balls are of the same radius $r$. If
$r^*_p < \alpha/2$, then $r = \alpha/2$ and $r = r^*_p$
otherwise.
Moreover, the leftmost
points of the balls (resp. rightmost points) are ordered in the same order as their centers, and thus the leftmost
point of balls in cluster $B_i$ is that of the ball
$CORE[i]$. It is easy to see that the cluster $B_i$ is
entirely contained in the ball of radius $4r$ centered
at the leftmost point of $CORE[i]$. The same holds for the
ball centered at its rightmost point. Thus the covering
constraints are satisfied.
In addition to this, the number of balls centered at the $c_i$ is $m\leq p$ and the number of balls centered at the $d_i$ is at most $p$ which is less than or equal to $q$. (To reach the exact numbers $p, q$ some balls can always be repeated.)

As for the factor of approximation and the separation constraint, we analyse two cases:

 \noindent \textbf{Case 1:}  $r^*_p\geq \alpha/2$: the distance between $c_i$ and $d_i$ is $2r^*_p$ which is at least $\alpha$. Moreover, we place the centers such that $d_{2k}$ and $d_{2k+1}$  are adjacent and $c_{2k}$ and $c_{2k+1}$  are adjacent. So for each two different color centers, we have $|c_i-d_j|\geq \alpha$. Note that the radius of balls is at most $4r^*_{p\land q, \alpha}$, since $r^*_p\leq r^*_{p\land q,\alpha}$.

\noindent \textbf{Case 2:}  $r^*_p<\alpha/2$: In this case, the radius of these expanded balls is $r=\alpha/2$.  The distance between $c_i$ and $d_i$ is $2r = \alpha$. Similar to Case 1, for all centers, we have $|c_i-d_j|\geq  \alpha$. Moreover, the radius of circles centered at $c_i$ and $d_j$ is at most $4r^*_{p\land  q,\alpha}$ since the radii are $2\alpha$ and $\alpha\leq 2r^*_{p\land q,\alpha}$.

\noindent \textbf{Analysis.}
Notice that the $p$-center optimal solution can be
constructed in $O(n\log^2 n)$ time via the algorithm
of \cite{brass}. The ordering of the balls can then be
done in time $O(n\log n)$ by sorting their centers. Then, once $CORE[i]$ is known for some $i$, the remaining balls
in the cluster can be found by doing a predecessor query.
The total time required will be $O(n \log n)$. Thus we have the following theorem,
\begin{theorem}
\thmlab{constrained-approx-main}%
There is an algorithm for the constrained $(n, p \land q, \alpha)$ problem that runs in $O(n \log^2 n)$ time and
outputs a feasible solution with covering radius at most
$4\ropt$, where $\ropt$ is the optimal radius.
\end{theorem}

Notice that the approximation factor can be improved
to $(1+\eps)$, for any $\eps > 0$ by combining the above
result with the procedure of \thmref{feasibility-main}, that decides feasibility of a given radius $r$. This can be done by searching in the range $[\frac{R}{4}, R]$ where $R$ is the covering radius of the solution output by \thmref{constrained-approx-main}, at the finite set of points
of the form $\frac{R}{4}(1+\eps)^t$ for $0 \leq t \leq \ceil{\log_{1+\eps} 4}$, using say binary search. Since this
is quite standard we omit the details.

\section{Solving the equations to determine candidate values for \texorpdfstring{$r^*$}{r*}}
\apndxlab{eqsolve}%
Fix a value of $t \in \mathbb{Z}$ with $0 \leq t \leq p + q - 1$, and $i, k$ with $0 \leq i,k \leq n$. If $i \neq k$, then we have four possible equations that arise,
$a_i(r) = a_k(r) + t\alpha, b_i(r) = a_k(r) + t\alpha, a_i(r) = b_k(r) + t\alpha, b_i(r) = b_k(r) + t\alpha$ and if
$i = k$, then we only have one possible equation $b_i(r) = a_i(r) + t\alpha$. For the point $\pnt_i=(\pnt_{i1}, \ldots, \pnt_{id})$ we denote 
\[
d_i = \sqrt{\sum_{j=2}^d p_{ij}^2},
\]
for brevity.
For $i = k$, the equation $b_i(r) = a_i(r) + t\alpha$ can be easily seen to solve to the solution,
$r = \sqrt{t^2\alpha^2/4 + d_i^2}$. On the other hand, if $i \neq k$, then from $a_i(r)=b_k(r) + t\alpha$ and $b_i(r) = a_k(r) + t\alpha$, we have equations of the form,
\[
\sqrt{r^2 - d_i^2} + \sqrt{r^2 - d_k^2} = \gamma_{ik},
\]
where $\gamma_{ik} = p_{i1} - p_{k1} - t\alpha$, or $\gamma_{ik} = -p_{i1} + p_{k1} + t\alpha$.

Similarly from the equations $a_i(r) = a_k(r) + t\alpha$ and $b_i(r) = b_k(r) + t \alpha$,
we have the equations of the form,
\[
\sqrt{r^2 - d_i^2} - \sqrt{r^2 - d_k^2} = \beta_{ik},
\]
where $\beta_{ik} = p_{i1} - p_{k1} - t\alpha$, or, $\beta_{ik} = -p_{i1} + p_{k1} + t\alpha$. Both of these kinds of equations can be solved via the same method, so we show only how to solve the first kind. We note however, that for the second kind of equation, without the general position assumption which is that $d_i \neq d_k$ for $i \neq k$, we could have infinitely many solutions for example when
$d_i = d_k$ and $\beta_{ik} = 0$. Because of the general position assumption there can only be finitely many solutions.
To solve the first type of equation, notice
there are no solutions if $\gamma_{ik} < 0$. For
$\gamma_{ik} = 0$ there are no solutions under our general position assumption that $d_i \neq d_k$. So assume $\gamma_{ik} > 0$. Now we have the identity,
\[
\sqrt{r^2 - d_i^2} - \sqrt{r^2 - d_k^2} = \frac{(\sqrt{r^2 - d_i^2} - \sqrt{r^2 - d_k^2})(\sqrt{r^2 - d_i^2} + \sqrt{r^2 - d_k^2})}{\sqrt{r^2 - d_i^2} + \sqrt{r^2 - d_k^2}},
\]
which holds true since $$\sqrt{r^2 - d_i^2} + \sqrt{r^2 - d_k^2} = \gamma_{ik} > 0.$$ Therefore we can simplify and write,
\[
\sqrt{r^2 - d_i^2} - \sqrt{r^2 - d_k^2} = \frac{d_k^2 - d_i^2}{\gamma_{ik}}.
\]
Thus we have that,
\[
\sqrt{r^2 - d_i^2} = 1/2 \left( \gamma_{ik} + (d_k^2 - d_i^2)/\gamma_{ik} \right), \text{ and, } \sqrt{r^2 - d_k^2} = 1/2 \left( \gamma_{ik} - (d_k^2 - d_i^2)/\gamma_{ik} \right).
\]
Thus the equation has a solution if and only if,
$\gamma_{ik} \geq \abs{d_k^2 - d_i^2}/\gamma_{ik}$ and this can now be easily computed.

\section{Removing the general position assumption}
\apndxlab{nogp}%
Our general position assumption states that for any two points 
$\pnt_i, \pnt_{i'} \in \PntSet, i \neq i'$,
their distance
from $\Line$ are not equal. Notice
that this was not required in the 
proof of \lemref{candidate}. However, it was required in the proof of \lemref{finite-sol}, and therefore ultimately in the proof of our main result \thmref{constrained-main}. To see this, suppose we have
some pair of indices $1 \leq i, i' \leq n, i \neq i'$ such that, $\sum_{j=2}^d \pnt_{ij}^2 = \sum_{j=2}^d \pnt_{i'j}^2$, and $\pnt_{ij} - \pnt_{i'j} = t\alpha$ for some $0 \leq t < p + q$. Then it follows that,
$\lept_i(r) - \lept_{i'}(r) = t\alpha$ for all $r$ and similarly,
$\rept_i(r) - \rept_{i'}(r) = t\alpha$ for all $r$. This leads to infinitely many possible $r$ as solutions to the equations 
$\lept_i(r) - \lept_{i'}(r) = t\alpha$ or $\rept_i(r) - \rept_{i'}(r) = t\alpha$ and thus \lemref{finite-sol} would not be valid, as it considers the union of all solutions as candidates for $r^*$. Here we show that in fact we do not need to consider equations from such pairs at all. We need a slight change in the statement of \lemref{candidate} and the corresponding proof.


First we give a definition. A pair $(i,i')$ is called \emphi{exceptional} if points $\pnt_i$ and $\pnt_{i'}$ are at the same distance from $\Line$ and for some integer $0 \leq t<p+q$, $\pnt_{i1}-\pnt_{i'1}=t\alpha$.
So, for an exceptional pair $(i,i')$, we have 
$\lept_{i}(r)-\lept_{i'}(r)=t \alpha$
and $\rept_i(r)-\rept_{i'}(r)=t\alpha$ for all values of $r$. We will say that a \emphi{distance between endpoints is exceptional} if it arises as the distance between the left (resp. right) endpoints of an exceptional pair. Otherwise such a 
distance is \emphi{non exceptional}.
The more general statement we need for \lemref{candidate} is as follows.
\begin{lemma}[Generalization of \lemref{candidate}]
 Let $U=\{u_1, u_2, \dots, u_{p+q}\}$ be a feasible solution for the constrained $(n,p\land q,\alpha)$ problem with  radius of covering $r$. If any non exceptional distance between two endpoints of the intervals in $\CandSet(r)$ is not
 $t \alpha$, where $t\in \mathbb{Z}, 0\leq t\leq p+q-1$,
 then the constrained $(n,p\land q,\alpha)$ problem has a feasible solution with radius less than  $r$.
 \lemlab{general}
\end{lemma}
\begin{proof}
We show how to deal differently with exceptional distances compared
to the proof of \lemref{candidate}.
Suppose that none of the non exceptional distances equal 
$t\alpha$ for some $0 \leq t < p + q$.
After converting the feasible centers to standard form, we find all alternate sequences, as we did in the proof of \lemref{candidate}. For each exceptional pair $(i,i')$, if none of the centers lie on the points $a_{i'}(r), a_{i}(r), b_i(r), b_{i'}(r)$, all the argumentation of the proof of  \lemref{candidate} is still valid and we can decrease $r$ in a similar way. So assume some centers do lie on such endpoints.
Here, one point needs to be made precise. When doing a perturbation $r \to r - \epsilon$, we move centers on endpoints with the endpoints themselves. There was no problem 
in \lemref{candidate} since we assumed all endpoints are distinct, i.e., no solutions exist for $t = 0$. Here, we are assuming only the non exceptional distances are not $t \alpha$ for $0 \leq t < p + q$. However, there could be an exceptional distance 
that is 0, i.e., there may be some exceptional pair $(i,i')$ is such that $a_i(r) = a_{i'}(r)$, and also in this case $b_i(r) = b_{i'}(r)$.
Now, if a center coincides with such an endpoint it is simultaneously on two endpoints. However, its movement is still determined without ambiguity, since these two endpoints always stay together for all $r$.

Now, let $S(k,j)$ be an alternate sequence such that $u_k=a_{i'}(r)$ and $u_{k+j}=a_{i}(r)$ and $a_{i'}(r)-a_{i}(r)=j \alpha$ for an exceptional pair $(i,i')$. For this exceptional pair we clearly may assume that $p_{i1} - p_{i'1} \neq 0$ since otherwise the endpoints would be the exact same point and this would not be an alternate sequence.

For constructing $\bar{U}$, we add an extra condition compared to proof of \lemref{candidate}: if $u_z$ is an internal center of such $S(k,j)$, let $\bar{u}_z=u_z+(a_{i}(r-\epsilon)-a_{i}(r))$. This causes $\bar{u}_{z+1}-\bar{u}_z=u_{z+1}-u_z=\alpha$.
 We calculate $\delta_1$ and $\delta_2$ as in the proof of \lemref{candidate}, but for $\delta_3$, we do not consider endpoints of such sequences, because $\bar{u}_{k+j}-\bar{u}_{k+j-1}=\alpha$ and $\bar{u}_{k}-\bar{u}_{k-1}=\alpha$.
 Therefore, $\bar{U}$ is a feasible solution with the covering radius $r-\epsilon$.
\end{proof}

By \lemref{general}, we only need to figure out non exceptional distances, consider the resulting equations, and solve them as in \apndxref{eqsolve}. This leads to  a finite number of solutions to $r^*$ and thus \lemref{finite-sol} is still true.
\end{document}